\documentclass[10pt]{article}

\usepackage{graphics}
\usepackage{graphicx}
\usepackage{epsfig}
\usepackage{subfigure}

\usepackage{verbatim}
\usepackage{amsmath}
\usepackage{amssymb}
\usepackage{color}
\usepackage{bm}  
\usepackage{hyperref}
 \usepackage{stackrel}
 \usepackage{upgreek}
 \usepackage{csquotes}
 \usepackage{fullpage}
 \usepackage{cite}
 \def\eps{\varepsilon}

 \def\epso{\eps_{\scriptscriptstyle 0}}
\def\lambdao{\lambda_{\scriptscriptstyle 0}}
\def\muo{\mu_{\scriptscriptstyle 0}}
\def\ko{k_{\scriptscriptstyle 0}}

\def\##1{{\bf #1}}
\def\=#1{\underline{\underline #1}}
\def\^#1{\breve{#1}}
\def\`#1{{#1^\prime}}
\def\:#1{#1^{\prime\prime}}

\def\les{\left[}
\def\ris{\right]}
\def\lec{\left\{}
\def\ric{\right\}}

\def\epsa{\eps_{\rm a}}
\def\epsb{\eps_{\rm b}}
\def\epsc{\eps_{\rm c}}

\def\.{\mbox{ \tiny{$^\bullet$} }}

\def\ux{\hat{\#u}_{\rm x}}
\def\uy{\hat{\#u}_{\rm y}}
\def\uz{\hat{\#u}_{\rm z}}

\def\as{a_{\rm s}}
\def\ap{a_{\rm p}}
\def\rs{r_{\rm s}}
\def\rp{r_{\rm p}}
\def\ts{t_{\rm s}}
\def\tp{t_{\rm p}}

\def\rss{r_{\rm ss}}
\def\rsp{r_{\rm sp}}
\def\rps{r_{\rm ps}}
\def\rpp{r_{\rm pp}}
\def\tss{t_{\rm ss}}
\def\tsp{t_{\rm sp}}
\def\tps{t_{\rm ps}}
\def\tpp{t_{\rm pp}}

\def\Rss{R_{\rm ss}}
\def\Rsp{R_{\rm sp}}
\def\Rps{R_{\rm ps}}
\def\Rpp{R_{\rm pp}}
\def\Tss{T_{\rm ss}}
\def\Tsp{T_{\rm sp}}
\def\Tps{T_{\rm ps}}
\def\Tpp{T_{\rm pp}}

\def\aL{a_{\rm L}}
\def\aR{a_{\rm R}}
\def\rL{r_{\rm L}}
\def\rR{r_{\rm R}}
\def\tL{t_{\rm L}}
\def\tR{t_{\rm R}}

\def\rLL{r_{\rm LL}}
\def\rLR{r_{\rm LR}}
\def\rRL{r_{\rm RL}}
\def\rRR{r_{\rm RR}}
\def\tLL{t_{\rm LL}}
\def\tLR{t_{\rm LR}}
\def\tRL{t_{\rm RL}}
\def\tRR{t_{\rm RR}}

\def\RLL{R_{\rm LL}}
\def\RLR{R_{\rm LR}}
\def\RRL{R_{\rm RL}}
\def\RRR{R_{\rm RR}}
\def\TLL{T_{\rm LL}}
\def\TLR{T_{\rm LR}}
\def\TRL{T_{\rm RL}}
\def\TRR{T_{\rm RR}}

\def\As{A_{\rm s}}
\def\Ap{A_{\rm p}}
\def\AL{A_{\rm L}}
\def\AR{A_{\rm R}}

\def\Ts{T_{\rm s}}
\def\Tp{T_{\rm p}}
\def\TL{T_{\rm L}}
\def\TR{T_{\rm R}}
\def\Rs{R_{\rm s}}
\def\Rp{R_{\rm p}}
\def\RL{R_{\rm L}}
\def\RR{R_{\rm R}}

\def\thetainc{\theta_{\rm inc}}

\def\lambdaoBr{\lambda_{\scriptscriptstyle 0}^{\rm Br}}

\def\tLD{\mathcal{LD}_{\rm tru}}
\def\aLD{\mathcal{LD}_{\rm app}}
\def\tCD{\mathcal{CD}_{\rm tru}}
\def\aCD{\mathcal{CD}_{\rm app}}

\begin{document}
 
 \begin{center}
 
\textbf{Experimental and theoretical investigation of the
co-occurrence of linear and circular dichroisms for oblique
incidence of light on chiral sculptured thin films}\\
{Patrick D. McAtee}  and {Akhlesh Lakhtakia}\\
{NanoMM---Nanoengineered Metamaterials Group,
Department of Engineering Science and Mechanics, The Pennsylvania State University, University Park, Pennsylvania 16802, USA}
\end{center}

\begin{abstract}
Theory shows that a slab of a  dielectric structurally chiral material (DSCM) exhibits both linear and circular dichroisms because of its anisotropy and structural chirality, for normal as well as oblique incidence. This conclusion was confirmed by fabricating a chiral sculptured thin film and measuring the spectra of its reflectances and transmittances, both linear and circular. Signatures of the circular Bragg phenomenon are evident in the spectra of all reflectances, transmittances, absorptances, and dichroisms. Reversal of the structural handedness  of a DSCM and 
rotation of the  projection
of the direction
of propagation of the incident light  clockwise
instead of counterclockwise  about the axis of helicoidal nonhomogeneity  
simultaneously
changes the sign of  circular dichroism but has no effect on  linear dichroism.

\end{abstract}

\section{Introduction}

Free space is unirefringent as are isotropic dielectric materials that are the staple of undergraduate textbooks in the field of electromagnetics \cite{Harrington,Guru,Iskander}. In a
homogeneous unirefringent medium at a specific frequency, two plane waves of mutually orthogonal polarization states can propagate in a given direction, but both have the same wavenumber. In other words, their phase speeds and attenuation rates are identical but their polarization states are not.

Birefringent materials, often exemplified in undergraduate textbooks \cite{Ghatak,Hecht} by uniaxial dielectric materials such as calcite, are different from unirefringent mediums. Two plane waves of mutually orthogonal polarization states can propagate in a given direction in a homogeneous
birefringent material at a specific frequency, but with different wavenumbers. Thus, their phase speeds, attenuation rates, and polarization states are different.

One of the two plane waves propagating in a particular direction in a homogeneous birefringent material will have a higher attenuation rate than the other, leading to the phenomenon of dichroism. If both plane waves are linearly polarized, as happens in uniaxial and biaxial dielectric materials lacking gyrotropy \cite[Sec.~14.6.1]{Born}, the phenomenon is qualified as \textit{linear dichroism}. If both plane waves are circularly polarized, as happens in isotropic chiral materials \cite{Charney,Beltramibook},  the phenomenon is qualified as \textit{circular dichroism}. 

Both uniaxiality and gyrotropy are present in a homogeneous dielectric material such as a magnetoplasma or a homogeneous magnetic material such as a ferrite.
The relative  permeability dyadic of a magnetoplasma  is a scalar dyadic, whereas
the relative permittivity dyadic   can be stated in dyadic notation as \cite[Chap.~7]{Chen}
\begin{equation}
\=\eps_{\rm r}=
\epsa\uz\uz+\epsb\left(\ux\ux+\uy\uy\right)+\eps_{\rm g}\times\=I\,,
\end{equation}
where the relative permittivity scalars
$\epsa$ and $\epsb$  together
indicate uniaxiality; $\eps_{\rm g}\times\=I$ is the gyrotropic term with $\=I$ denoting
the identity dyadic;
and $\ux$, $\uy$, and $\uz$ are the three
Cartesian unit vectors.
The forms of the relative permeability and relative permittivity dyadics
are reversed for a ferrite \cite[Chap.~7]{Chen}.
Both linear dichroism and circular dichroism can  simultaneously occur in such a material, but circular dichroism alone is observable when both plane waves propagate parallel to its sole optic axis
(which is parallel to the $z$ axis). Circular dichroism can be induced in a unirefringent material
by immersion in a strong magnetic field oriented parallel to the direction of light propagation \cite{Caldwell,Stephens}.

Both linear dichroism and circular dichroism can simultaneous occur also in a non-gyrotropic dielectric material, provided it is structurally chiral \cite{Reusch,Bose,FLaop}. This hypothesis underlaid a theoretical investigation of the spectral signatures of an axially excited slab of a dielectric structurally chiral material (DSCM) \cite{Lakh1999}.
DSCMs are exemplified by chiral liquid crystals   \cite{deG,Chan,Nityananda} and chiral sculptured thin films (CSTFs) \cite{YK1959,STFbook,Erten2015}. The relative permittivity dyadic $\=\eps_{\rm r}$
of a DSCM is either uniaxial or biaxial and it varies helicoidally along a fixed axis (taken to be the $z$ axis in this paper). Thus
 \cite{FLaop},
 \begin{eqnarray}
&&\=\eps_{\rm r}^{\rm DSCM}=
\=S_{\rm z}(h,z,\Omega) \.
S_{\rm y}(\chi)\.\left[\epsa\uz\uz+\epsb\ux\ux
+\epsc\uy\uy\right]
\.\=S_{\rm y}^{-1}(\chi) \.\=S_{\rm z}^{-1}(h,z,\Omega)\,,
\end{eqnarray}
where the relative permittivity scalars
$\epsa$, $\epsb$, and $\epsc$ indicate the local orthorhombicity of the DSCM;
\begin{eqnarray}
 &&
\=S_{\rm y} (\chi) = \uy\uy + \left(\ux\ux+\uz\uz\right)\cos\chi
 + \left(\uz\ux-\ux\uz\right)\sin\chi
 \label{SY-def}
 \end{eqnarray}
contains $\chi\in(0\ \text{deg},90\ \text{deg}]$ is an angle of inclination with respect to the $xy$ plane;
 and the rotation dyadic
\begin{eqnarray}
&&
\=S_z(h,z,\Omega)=\uz\uz +\left(\ux\ux+\uy\uy\right)\cos(\pi{z}/\Omega)
+h\left(\uy\ux-\ux\uy\right)\sin(\pi{z}/\Omega)
\label{Sz-def}
\end{eqnarray} contains
$h \in\left\{-1,1\right\}$ as the structural
handedness parameter, where $h=-1$ for structural left-handedness
and $h=1$ for structural right-handedness, and $2\Omega$ as the helicoidal period
along the the $z$ axis. Gyrotropy has thus been replaced by helicoidal nonhomogeneity.

Theoretical investigation confirmed that when light is normally incident on a DSCM confined
between the planes $z=0$ and $z=L$, and the ratio $L/2\Omega$ is large enough,
the DSCM must exhibit the circular Bragg phenomenon \cite{FLaop}. The signature
of the circular Bragg phenomenon is twofold:
(i) almost total reflection of the incident light of the co-handed circular-polarization
state and (ii) very little reflection of the incident light of the cross-handed
circular-polarization state. This signature occurs in a spectral regime called the circular
Bragg regime.

The circular Bragg phenomenon was found to be accompanied by distinct features
in the spectrums of both linear and circular dichroisms \cite{Lakh1999}. 
Linear dichroism pertains
to the differential in the response to incident light of the $s$- and $p$-polarization states,
whereas circular dichroism pertains
to the differential in the response to incident light of the left-  and  right-circular polarization states.

Furthermore, a distinction emerged
between true and apparent dichroisms. True dichroism is the differential in the
absorptance whereas apparent dichroism is the differential in transmittance,
with apparent dichroism requiring less effort to measure than true dichroism. 
Where true dichroism can only occur in a dissipative material, apparent dichroism
can be exhibited even if dissipation is small enough to be ignored.

All data reported  by Lakhtakia\cite{Lakh1999} were calculated by assuming $\epsa=\epsc$ and
$\epsb$  to be complex as well as frequency dependent. Later, similar data were published with respect to the porosity of a CSTF \cite{SherwinMCM1,SherwinMCM2}.  Very recently,
nonzero values of true dichroisms of a zirconium-oxide CSTF  were reported
 for oblique incidence \cite{ZahirOM1,ZahirOM2}, but the data must be incorrect
 because all three of $\epsa$, $\epsb$, and $\epsc$ were assumed real and
positive  \cite{HWH}; moreover, $\epsa$, $\epsb$, and $\epsc$ were taken to be independent of frequency.

No experimental measurement of any of the four types of dichroisms---true linear
$\tLD$, true circular $\tCD$, apparent
linear $\aLD$, and apparent circular $\aCD$---have ever been reported, which motivated us to experimentally investigate
the co-exhibition of linear dichroism and circular dichroism by a DSCM.  For this purpose, we fabricated
a CSTF made of zinc selenide (ZnSe) using the serial-bideposition   technique \cite{Hodgkinson,McAtee}.
Spectral measurements of reflectance and transmittances were carried out for linearly as well as circularly polarized
 light obliquely incident of the CSTF over the 500-to-900-nm range of the free-space wavelength
 $\lambdao$
 \cite{Erten2015,McAtee,Vepachedu}. We also performed a theoretical investigation in parallel.

This paper is organized as follows. Section~2.\ref{sec:theory} provides 
an overview of the formalism to define and calculate $\tLD$,  $\tCD$,  
$\aLD$, and  $\aCD$.
An $ \exp(-i \omega t)$  dependence on time $t$ is explicit, with $ \omega$ as the angular frequency
and $ i = \sqrt{-1}$. The free-space wavenumber is denoted by $\ko = \omega \sqrt{\epso \muo }
=2 \pi / \lambdao$, with $ \epso$ and $\muo$
being the permittivity and permeability, respectively, of free space. Section~2.\ref{sec:expt} contains
a description of the experimental procedures used for the fabrication as well as  
the morphological and optical characterization of the CSTF. Theoretical and experimental data for $\tLD$,  $\tCD$,  
$\aLD$, and  $\aCD$ are presented and discussed in Secs.~\ref{theoresults} and \ref{exptresults}, respectively.   


\section{Methods}

\subsection{\label{sec:theory}Theory}
 
\subsubsection{Linear Reflectances and Transmittances}

Suppose the region $ 0 < z < L $ is occupied by a CSTF of infinite transverse extent. The electric field phasor of the plane wave incident on the CSTF from the half space $ z<0 $ is given by \cite{STFbook,Vepachedu}
\begin{eqnarray}
&&
 \#E_{\rm inc}(x,y,z)  = \left(\as  \hat{\#s} + \ap \hat{\#p}_+\right)
 \exp\left\{i   {\ko}  \les\left({x\cos\psi+y\sin\psi}\right)   \times \sin\thetainc +z   \cos \thetainc\ris
 \ric\,,\quad z< 0\,,
 \end{eqnarray} 
 where $ \thetainc \in [0\ \text{deg}, 90\ \text{deg}) $ is the angle of incidence with respect to the
 $z$ axis, $\psi\in[0\ \text{deg},360\ \text{deg})$ is the angle of incidence with respect to the
 $x$ axis in the $xy$ plane, $\as$ is the amplitude of the $s$-polarized component,
  and $\ap$ is the amplitude of the $p$-polarized component. Here and hereafter, the unit vectors  
 \begin{equation}
 \left.\begin{array}{l}
\hat{\#s} = -\ux \sin\psi + \uy \cos\psi
\\[5pt]
\hat{\#p}_{\pm} = {\mp}(\ux \cos\psi + \uy \sin\psi)\cos\thetainc + \uz \sin\thetainc 
\end{array}\right\}
\end{equation}
delineate the orientations of the various field phasors.

The electric field phasor of the reflected plane wave is given by 
\begin{eqnarray}
&&
 \#E_{\rm ref}(x,y,z)  = \left(\rs  \hat{\#s} + \rp \hat{\#p}_-\right)
 \exp\left\{i   {\ko}  \les\left({x\cos\psi+y\sin\psi}\right)   
\sin\thetainc -z   \cos \thetainc\ris
 \ric\,,\quad z< 0\,,
 \end{eqnarray} and the electric
field phasor of the transmitted plane wave by  
\begin{eqnarray}
&&
 \#E_{\rm tr}(x,y,z)  = \left(\ts  \hat{\#s} + \tp \hat{\#p}_+\right)
 \exp\left\{i   {\ko}  \les\left({x\cos\psi+y\sin\psi}\right)    \times \sin\thetainc -(z-L)   \cos \thetainc\ris
 \ric\,,\quad z>L\, .
 \end{eqnarray}
 
 The reflection amplitudes $\rs$ and $\rp$, as well as the transmission
amplitudes $\ts$ and $\tp$, have to be determined in terms of the
incidence amplitudes $\as$ and $\ap$. The procedure is described in detail
elsewhere \cite[Chap.~9]{STFbook}. Thereafter,
we can calculate 
the eight linear reflection and transmission coefficients appearing as the elements of the 
2$\times$2 matrices in the following
relations \cite{STFbook,Vepachedu}:
\begin{eqnarray}
\label{eq10}
\les\begin{array}{c}\rs\\\rp\end{array}\ris
=
\les\begin{array}{cc}\rss & \rsp\\\rps & \rpp\end{array}\ris
\,
\les\begin{array}{c}\as\\\ap\end{array}\ris\,,\\[5pt]
\les\begin{array}{c}\ts\\\tp\end{array}\ris
=
\les\begin{array}{cc}\tss & \tsp\\\tps & \tpp\end{array}\ris
\,
\les\begin{array}{c}\as\\\ap\end{array}\ris\,.
\end{eqnarray}
Co-polarized coefficients have both subscripts identical, but
cross-polarized coefficients do not. The square of the magnitude
of a linear reflection or transmission coefficient is the corresponding linear
reflectance or transmittance;  thus, $\Rsp = \vert\rsp\vert^2$ is
the linear reflectance corresponding to the linear reflection coefficient $\rsp$,
and so on. 

\subsubsection{Circular Reflectances and Transmittances}

The procedure to determine the circular reflectances and transmittances
is very similar \cite[Chap.~9]{STFbook}. Let
$\aL$ be the amplitude of the left-circularly polarized component
and $\aR$ be the amplitude of the right-circularly polarized component
 of the incident plane wave. Then,  
\begin{equation}
\label{eq11}
\left.\begin{array}{l}
\aL = -\frac{1}{\sqrt{2}} (i \as+\ap )
\\[5pt]
\aR = \frac{1}{\sqrt{2}} ( i\as-\ap )
\end{array}\right\}\,.
\end{equation}
With $\rL$, $\rR$, $\tL$, and $\tR$ denoting the circularly polarized
components of the reflected and transmitted plane waves, we have
\begin{equation}
\label{eq11}
\left.\begin{array}{l}
\rL = \frac{1}{\sqrt{2}} (i \rs+\rp )
\\[5pt]
\rR =- \frac{1}{\sqrt{2}} ( i\rs-\rp )
\end{array}\right\}\,.
\end{equation}
and
\begin{equation}
\label{eq11}
\left.\begin{array}{l}
\tL = -\frac{1}{\sqrt{2}} (i \ts+\tp )
\\[5pt]
\tR = \frac{1}{\sqrt{2}} ( i\ts-\tp )
\end{array}\right\}\,.
\end{equation}

Accordingly, 
 eight circular reflection and transmission coefficients can be defined as the elements of the 
2$\times$2 matrices in the following
relations \cite{STFbook}:
\begin{eqnarray}
\label{eq10}
\les\begin{array}{c}\rL\\\rR\end{array}\ris
=
\les\begin{array}{cc}\rLL & \rLR\\\rRL & \rRR\end{array}\ris
\,
\les\begin{array}{c}\aL\\\aR\end{array}\ris\,,
\\[5pt]
\les\begin{array}{c}\tL\\\tR\end{array}\ris
=
\les\begin{array}{cc}\tLL & \tLR\\\tRL & \tRR\end{array}\ris
\,
\les\begin{array}{c}\aL\\\aR\end{array}\ris\,.
\end{eqnarray}
The square of the magnitude
of a circular reflection or transmission coefficient is the corresponding circular
reflectance or transmittance;  thus, $\RLR = \vert\rLR\vert^2$ is
the circular reflectance corresponding to the circular reflection coefficient $\rLR$,
and so on.

\subsubsection{True Dichroisms}
\label{true-dichroism}
True dichroisms quantify   differential absorption. Hence, we define the linear absorptances
\begin{eqnarray}
\left.\begin{array}{l}
 \As = 1 - \left(\Rs  + \Ts\right)\in[0,1]
 \\[5pt]
\Ap = 1 - \left(\Rp + \Tp\right)\in[0,1]
 \end{array}\right\}\,
 \end{eqnarray} 
 and
 the circular absorptances
 \begin{eqnarray}
\left.\begin{array}{l}
\AL = 1 - \left(\RL  + \TL \right)\in[0,1]
\\[5pt]
\AR = 1 - \left(\RR +  \TR\right)\in[0,1]
 \end{array}\right\}\,,
 \end{eqnarray} 
 where the total reflectances
 \begin{equation}
 \left.\begin{array}{l}
 \Rs=\Rss+\Rps\in[0,1]
 \\[5pt]
 \Rp=\Rpp+\Rsp\in[0,1]
 \\[5pt]
  \RL=\RLL+\RRL\in[0,1]
   \\[5pt]
  \RR=\RRR+\RLR\in[0,1]
\end{array}\right\}\,
\end{equation}
and the total transmittances
 \begin{equation}
 \left.\begin{array}{l}
 \Ts=\Tss+\Tps\in[0,1]
 \\[5pt]
 \Tp=\Tpp+\Tsp\in[0,1]
 \\[5pt]
  \TL=\TLL+\TRL\in[0,1]
   \\[5pt]
  \TR=\TRR+\TLR\in[0,1]
\end{array}\right\}\,.
\end{equation}

The true linear dichroism is then given by 
\begin{equation}
\tLD = \sqrt{\As} - \sqrt{\Ap}
\label{tLD-def}
\end{equation}
and the true circular dichroism by  
\begin{equation}
\tCD = \sqrt{\AR} - \sqrt{\AL}\,.
\label{tCD-def}
\end{equation}

\subsubsection{Apparent Dichroisms}
\label{apparent-dichroism}
Apparent dichroisms quantify   differential transmission. Hence, the apparent linear dichroism is  given by \cite[Chap. 9]{STFbook}
\begin{equation}
\aLD = \sqrt{\Ts} - \sqrt{\Tp}
\label{aLD-def}
\end{equation}
and the apparent circular dichroism by \cite[Chap. 9]{STFbook}
\begin{equation}
\aCD = \sqrt{\TR} - \sqrt{\TL}\,.
\label{aCD-def}
\end{equation}

\subsubsection{Calculation Parameters}

All calculations were made by assuming single-resonance Lorentzian dependences on $\lambdao$
for $\epsa$, $\epsb$, and $\epsc$;
thus \cite{Kittel},
\begin{equation}
\label{resonance}
\eps_{\rm a,b,c}(\lambdao) = 1+ \frac{p_{\rm a,b,c}}
{1 + (1/N_{\rm a,b,c}  - i \lambda_{\rm a,b,c}/ \lambdao )^2}\,.
\end{equation}
The oscillator strengths are determined by the values of $p_{\rm a,b,c}$,  $\lambda_{\rm a,b,c} (1 + N_{\rm a,b,c}^{-2})^{-1/2}  $  are the  resonance wavelengths, and $\lambda_{\rm a,b,c}/N_{\rm a,b,c}$ are the  resonance linewidths. The larger the values of $N_{\rm a,b,c}$, the narrower are the absorption bands.

 The parameters used for theoretical data reported here are as follows:  $p_{\rm a} = 2.3$, $p_{\rm b} =3.0$, $p_{\rm c} =2.2 $, $\lambda_{\rm a} = \lambda_{\rm c} =260$~nm, $\lambda_{\rm b} = 270$~nm,   and
$N_{\rm a} = N_{\rm b} =N_{\rm c}=130$. Furthermore, $h = 1$, $\chi = 37$~deg, $\Omega = 150$~nm,  $L = 20\Omega$, and $\psi=0$~deg.

\subsection{Experiments}\label{sec:expt}

\subsubsection{Fabrication of CSTF}
\label{fabrication}

The CSTF for this project was fabricated using thermal evaporation implemented inside a low-pressure chamber \cite{Swiontek_1} from Torr International (New Windsor, New York). It contains a quartz crystal monitor  calibrated to measure the growing thin-film's thickness, a receptacle to hold the material to be evaporated to generate a collimated vapor flux, electrodes to resistively heat the receptacle, and a substrate holder positioned about 15 cm directly above the receptacle. As the chamber was customized to grow CSTFs \cite{Erten2015}, it contains two stepper motors to control the rotation of the substrate holder about two mutually orthogonal axes, one passing normally through the substrate holder to serve as the $z$ axis, and the second 
serving as the $y$ axis in the substrate ($xy$) plane.

 The motors were programmed such that $\Omega = 150$~nm. The angle of the collimated vapor flux with 
respect to the substrate plane set at $\chi_v = 20$~deg corresponding to $\chi=37$~deg, as determined
experimentally elsewhere \cite{McAtee}. 99.995 \% pure ZnSe (Alfa Aesar, Ward Hill, Massachusetts) was the material of choice due to its high bulk index of refraction and low absorption in the visible spectral regime \cite{ThorLabs,Tydex} as well as its ease of evaporation. The manufacturer supplied ZnSe lumps that were crushed into a fine powder. A respirator was worn to avoid the toxic effects of ZnSe on the respiratory system \cite{Alfa}. 

Approximately 4.2 g of ZnSe powder was packed into a tungsten boat (S22-.005W, R. D. Mathis, Long Beach, California) that served as the receptacle.
Then the substrate was 
 further cleaned in an ethanol bath using an ultrasonicator for 10~min on each side. After removal from the bath, the substrate was immediately dried with pressurized nitrogen gas. The substrate was carefully secured to the substrate holder  using Kapton tape (S-14532, Uline, Pleasant Prairie, Wisconsin), and a shutter was interposed between the receptacle and the substrate holder. By the side of the glass substrate, a silicon wafer was positioned in order to grow a sample for imaging on a scanning electron microscope.

The chamber was pumped down to 1~$\mu$Torr. The current through the receptacle was then slowly increased to
$\sim$~100 A, and the shutter was rotated to allow a collimated portion of the ZnSe vapor flux to reach the substrate and the wafer. The deposition rate as read through the quartz crystal monitor  was manually maintained equal to $0.4 \pm 0.02$~nm~s$^{-1}$. After the deposition was completed, the shutter was rotated to prevent further deposition, the current was brought back to 0 A, the chamber was allowed to cool down for 60 min, and was then exposed to the atmosphere. The CSTF was grown in two stages, each depositing $5\times 2\Omega$. Thus, the total thickness was $L=20\Omega$. 

The CSTF was grown using the serial-bideposition technique described as follows \cite{Hodgkinson}. A subdeposit is made for time $\tau_1$, followed by a rapid rotation about the $z$ axis by angle $\Psi$  in time  $\tau_{\rm a}$, followed by another
subdeposit for time $\tau_1$, followed by a rapid  rotation about the $z$ axis by angle $\Psi+\delta$  in time  $\tau_{\rm b}>\tau_{\rm a}$, and so on. Therefore, the total time to deposit one pair of subdeposits is given by $\tau_{\rm sdp} = 2\tau_1 + \tau_{\rm a} + \tau_{\rm b}$. If $N_{\rm sdp}$ is the number of
subdeposit-pairs per period, the total deposition time for one period is given by $\tau_{\rm per} = N_{\rm sdp} \tau_{\rm sdp}$.  We fixed $\Psi=180$~deg, $\delta=3$~deg,
$\tau_{\rm  a} = 0.406$~s,  $\tau_{\rm b} = 0.413$~s, and $\tau_{\rm per} = 750$~s. Accordingly,
$N_{\rm sdp} = 360/3 = 120$, $\tau_{\rm sdp} = 750/120 = 6.25$~s, and $\tau_1 = 2.7155$~s.

\subsubsection{Morphological Characterization}

The cross-sectional morphology of the CSTF sample was determined using an FEI
Nova  NanoSEM 630 (FEI, Hillsboro, Oregon) field-emission scanning electron microscope.
To get a clear image of morphology away from any edge-growth effects, each sample was
cleaved using the freeze-fracture technique \cite{Severs}. The sample was sputtered with iridium using a Quorum Emitech K575X (Quorum Technologies, Ashford, Kent, United Kingdom) sputter coater before imaging.

\subsubsection{Optical Characterization}
\label{opt characterization}

All transmittance and reflectance measurements were made within 24~h after fabrication. The sample was kept in a desiccator up until the time of optical characterization in order to prevent degradation due to moisture adsorption.
The experimental setups for reflection and transmission measurements for
$\lambdao\in[500\ \text{nm},900\ \text{nm}]$ are described in detail elsewhere \cite{McAtee,Vepachedu}. All measurements were taken in a dark room to avoid noise from external sources,  and we ensured that $\psi$ was the same for all measurements.  

Briefly, light from a halogen source (HL-2000, Ocean Optics, Dunedin, Florida) was passed through a fiber-optic cable and then through a linear polarizer (GT10, ThorLabs, Newton, New Jersey); it was either reflected from or transmitted through the sample to be characterized and was then passed through a second linear polarizer (GT10, ThorLabs) and a fiber-optic cable to a CCD spectrometer (HRS-BD1-025, Mightex Systems, Pleasanton, California) \cite{Vepachedu}. The linear
transmittances $\Tss$, $\Tps$, $\Tsp$, and $\Tpp$ were measured for $\thetainc \in [ 0 \ \text{deg}, 70 \ \text{deg} ] $, and the linear reflectances $\Rss$, $\Rps$, $\Rsp$, and $\Rpp$  
  for $\thetainc \in [ 10 \ \text{deg}, 70 \ \text{deg} ]$.

For measurements of the
circular reflectances and transmittances, a Fresnel rhomb (LMR1, ThorLabs) was introduced directly after the first linear polarizer and another Fresnel rhomb
directly before the second linear polarizer \cite{McAtee}. The circular
transmittances $\TLL$, $\TRL$, $\TLR$, and $\TRR$ were measured for $\thetainc \in [ 0 \ \text{deg}, 70 \ \text{deg} ] $, and the linear reflectances $\RLL$, $\RRL$, $\RLR$, and $\RRR$  
  for $\thetainc \in [ 10 \ \text{deg}, 70 \ \text{deg} ]$.

\section{Theoretical Results}\label{theoresults}
\subsection{Circular Reflectances and Transmittances}
\begin{figure}
\begin{center}
\begin{tabular}{c}
\includegraphics[width=0.55\linewidth]{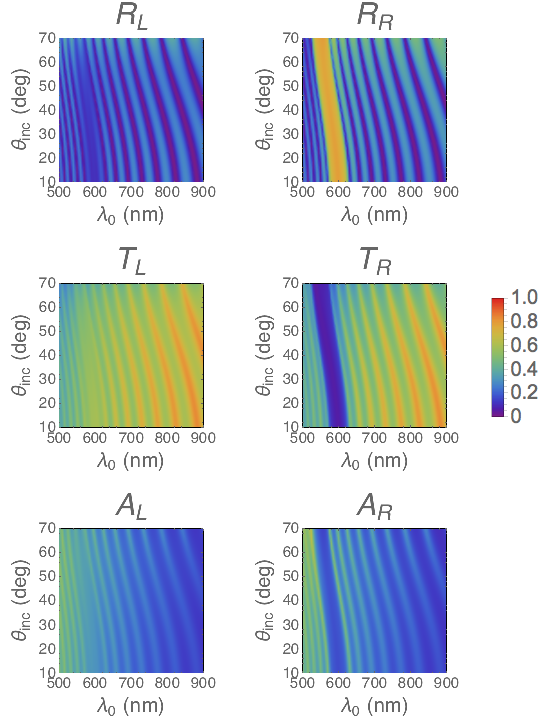}
\end{tabular}
\end{center}
\caption{\label{Theor-CircRemit}
$\RL$, $\RR$, $\TL$, $\TR$, $\AL$, and $\AR$ calculated as functions
of $\lambdao$ and $\thetainc$ for $h=1$ and $\psi=0$~deg. 
}
\end{figure}

As DSCMs exhibit the circular Bragg phenomenon, their circular reflectances and transmittances for normal incidence have been of major
interest for both theorists and experimentalists for several decades \cite{FLaop}. Theoretical results for oblique incidence
have also been published \cite{BS1970,OAT1985,ABW1985,VLprsa2000,SH2001,DLoc2008,Takezoe}, but experimental data for oblique incidence are rare and incomprehensive \cite{BS1970,Takezoe,Jacobs,StJohn,HWTLM,Podraza,vanPopta}. Although spectrums of the circular reflectances and transmittances over wide ranges of both $\lambdao$ and $\thetainc$ have recently been published \cite{Erten2015,Erten-jnp}, we provide calculated values of $\RL$, $\RR$, $\TL$, and $\TR$ as functions
of $\lambdao$ and $\thetainc$ in Fig.~\ref{Theor-CircRemit} both for completeness as well as to contextualize novel results.

The circular Bragg phenomenon is manifest in Fig.~\ref{Theor-CircRemit} as a high  ridge in the graph of $\RR$ and a correspondingly deep trough in the graph of $\TR$. 
For $\thetainc\in\left\{10\ \text{deg}, 30\ \text{deg},70\ \text{deg}\right\}$, values of the center wavelength
of
the circular Bragg regime were estimated  as  $\lambdaoBr\in\left\{599\ \text{nm}, 584\ \text{nm}, 549\ \text{nm}\right\}$ 
from both of these graphs. Thus, the circular Bragg regime blueshifts as the incidence becomes more oblique. A trough in
the graph of $\RL$ and a ridge in the graph of $\TL$ are also evident. 

All of the foregoing features arose because we set $h=1$ for calculations. As a result, the CSTF was chosen to be structurally right handed so that incident right-circularly polarized light must be reflected much more than  incident left-circularly polarized light in the circular Bragg regime. If we had set $h=-1$, the CSTF would have been structurally left handed so that incident left-circularly polarized light would have been reflected much more than  incident right-circularly polarized light.

\subsection{True and Apparent Circular Dichroisms}
Figure~\ref{Theor-CircRemit} also shows graphs of  $\AL$ and $\AR$ as functions
of $\lambdao$ and $\thetainc$. The circular Bragg phenomenon is faintly evident as a  ridge in the graph of $\AL$ and clearly as
a trough in the graph of $\AR$.   Therefore, in the circular Bragg regime, $\AL\ne\AR$ so that $\tCD\ne0$; furthermore,
$\aCD\ne0$ because  $\TL\ne\TR$ in Fig.~\ref{Theor-CircRemit}.

\begin{figure}
\begin{center}
\begin{tabular}{c}
\includegraphics[width=0.55\linewidth]{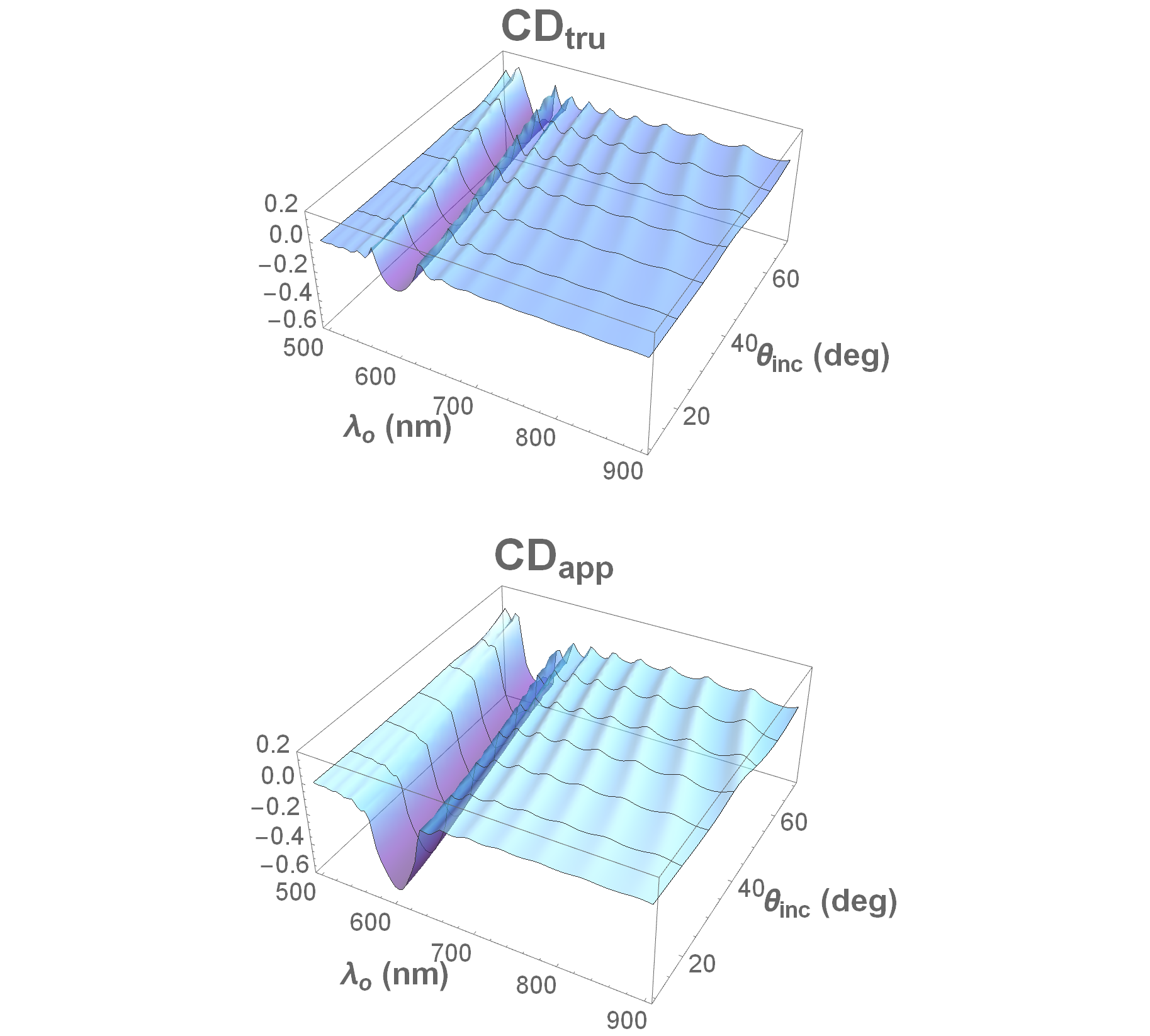}
\end{tabular}
\end{center}
\caption{\label{Theor-tCDaCD}
$\tCD$  and $\aCD$ calculated as functions
of $\lambdao$ and $\thetainc$ for $h=1$ and $\psi=0$~deg. 
}
\end{figure}

Graphs of $\tCD$ and $\aCD$ as functions of $\lambdao$ and $\thetainc$ are provided
in Fig.~\ref{Theor-tCDaCD}. Both graphs have troughs coincident with the manifestation
of the circular Bragg phenomenon in Fig.~\ref{Theor-CircRemit}. For $\thetainc\in\left\{10\ \text{deg}, 30\ \text{deg},70\ \text{deg}\right\}$, values of the center wavelength 
of
the circular Bragg regime were estimated  as  $\lambdaoBr\in\left\{598\ \text{nm}, 584\ \text{nm},546\ \text{nm}\right\}$ 
from  both graphs in Fig.~\ref{Theor-tCDaCD}. These estimates are just  $1.0$~nm, $0.0$~nm, and $3.0$~nm different from the estimates drawn from Fig.~\ref{Theor-CircRemit}.

Given the definition of $\tCD$ in Eq.~(\ref{tCD-def}), we expect that $\tCD<0$ in the circular Bragg regime because
$\lec{h=1,L\gg2\Omega}\ric\implies \AR<\AL$. Likewise, in view
of its definition  in Eq.~(\ref{aCD-def}), we expect that $\aCD<0$ in the circular Bragg regime because
$\lec{h=1,L\gg2\Omega}\ric\implies \TR<\TL$ if the CSTF is not strongly dissipative. 
Both of these expectations are borne out by the graphs in Fig.~\ref{Theor-tCDaCD}, especially in the central portion of the circular Bragg regime.

Let us also note a significant symmetry evinced by circular reflectances and transmittances. If the replacement $\lec{h\to-h,\psi\to-\psi}\ric$ is made, then the interchanges
$\lec{\RRR\leftrightarrow\RLL, \RLR\leftrightarrow\RRL, \TRR\leftrightarrow\TLL, \TLR\leftrightarrow\TRL}\ric$ follow. Accordingly, the replacement $\lec{h\to-h,\psi\to-\psi}\ric$
implies that $\lec{\tCD\to-\tCD, \aCD\to-\aCD}\ric$. For arbitrary values of $\psi$, the replacement $h\to-h$ affects both $\tCD$ and $\aCD$.

\subsection{Linear Reflectances and Transmittances}
\begin{figure}
\begin{center}
\begin{tabular}{c}
\includegraphics[width=0.55\linewidth]{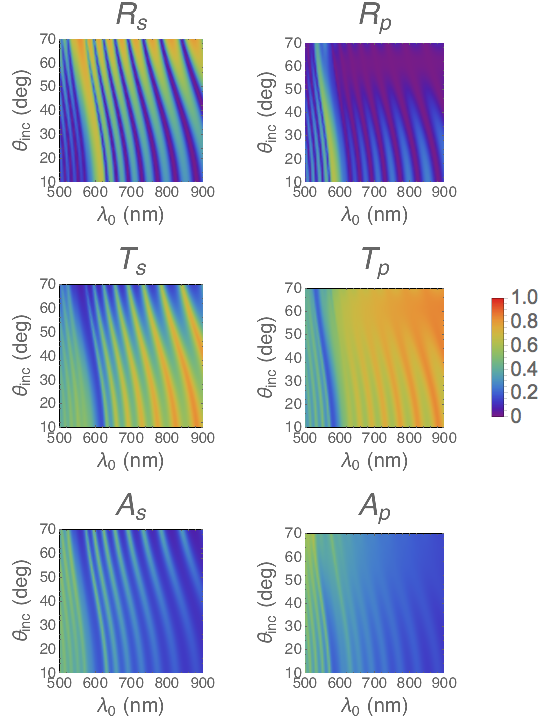}
\end{tabular}
\end{center}
\caption{\label{Theor-LinRemit}
 $\Rs$, $\Rp$, $\Ts$, $\Tp$, $\As$, and $\Ap$ calculated as functions
of $\lambdao$ and $\thetainc$ for $h=\pm1$ and $\psi=0$~deg. 
}
\end{figure}

The linear   reflectances and transmittances of DSCMs have been investigated very little \cite{Lakh1999,STFbook}, for which reason we provide calculated values of $\Rs$, $\Rp$, $\Ts$, and $\Tp$ in relation to both $\lambdao$ and $\thetainc$ in Fig.~\ref{Theor-LinRemit}. In these graphs, the circular Bragg phenomenon has a split signature: a  ridge and a trough side by side in the graphs of both $\Rs$ and $\Rp$,
and a trough  and a ridge side by side  in the graphs of both $\Ts$ and $\Tp$. The blueshift of the circular Bragg phenomenon with oblique incidence is also captured in these graphs.

\subsection{True and Apparent Linear Dichroisms}

\begin{figure}
\begin{center}
\begin{tabular}{c}
\includegraphics[width=0.55\linewidth]{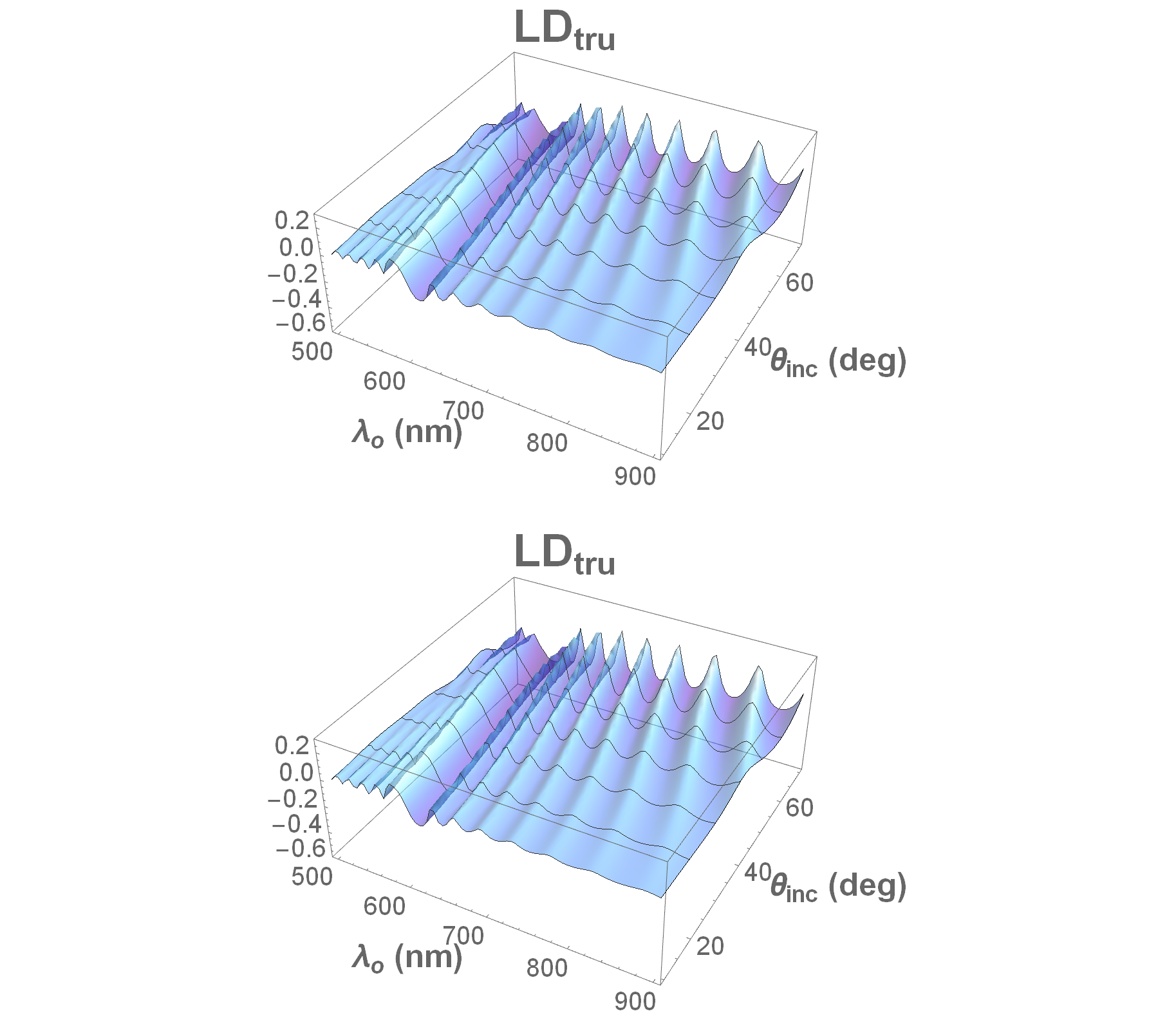}
\end{tabular}
\end{center}
\caption{\label{Theor-tLDaLD}
$\tLD$  and $\aLD$ calculated as functions
of $\lambdao$ and $\thetainc$ for $h=\pm1$ and $\psi=0$~deg.
}
\end{figure}

In general, $\Ts$ and $\Tp$ are unequal in Fig.~\ref{Theor-LinRemit} 
as also are $\As$ and $\Ap$. Therefore, $\tLD\ne0$ and $\aLD\ne0$, as depicted
in Fig.~\ref{Theor-tLDaLD}. The graphs of $\tLD$ and $\aLD$ as functions of
$\lambdao$ and $\thetainc$ clearly trace out the circular Bragg phenomenon in the form of a ridge
by the side of a trough, both of which blueshift as $\thetainc$ increases.

A change in the structural handedness does not affect the graphs in Fig.~\ref{Theor-LinRemit} because $\psi=0$~deg. Indeed, all eight linear reflectances and transmittances remain unaffected by the replacement $\lec{h\to-h,\psi\to-\psi}\ric$. Hence, that replacement  affects neither $\tLD$ nor $\aLD$.
For arbitrary values of $\psi$, the replacement $h\to-h$ affects both $\tLD$ and $\aLD$.

\section{Experimental Results}\label{exptresults}
\subsection{Morphology}

\begin{figure}
\begin{center}
\begin{tabular}{c}
\includegraphics[width=0.55\linewidth]{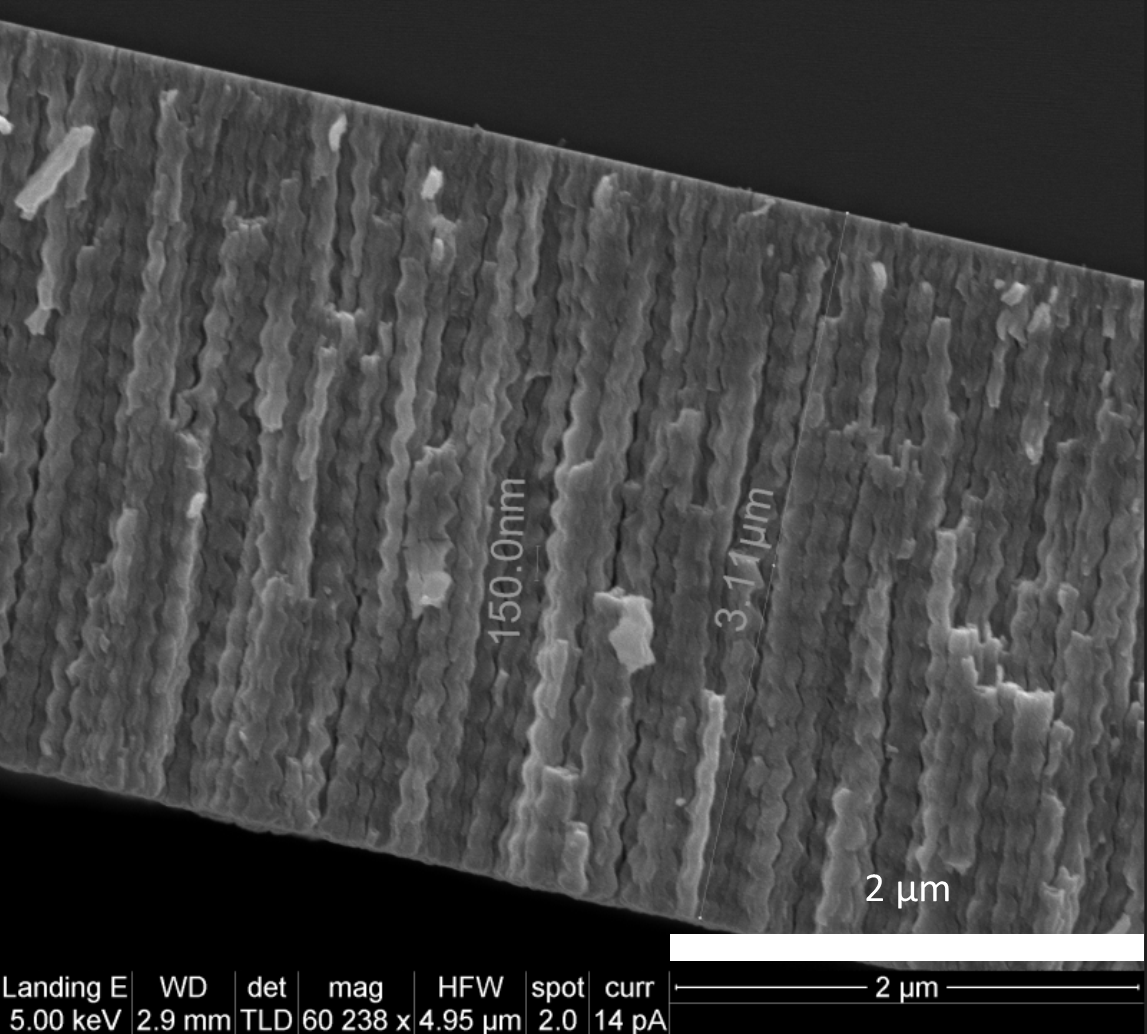}
\end{tabular}
\end{center}
\caption{\label{SEM-image}Cross-sectional scanning-electron micrograph of the ZnSe CSTF fabricated
for this research.
}
\end{figure}

The scanning-electron micrograph in Fig.~\ref{SEM-image} shows that the overall thickness $L$ of the ZnSe CSTF is $3.11$~$\mu$m; thus, the  period $2\Omega$ is equal to $311$~nm. This is just 3.7\% more than the targeted value of $300$~nm.

\subsection{Circular Reflectances, Transmittances, and Dichroisms}

\begin{figure}
\begin{center}
\begin{tabular}{c}
\includegraphics[width=0.55\linewidth]{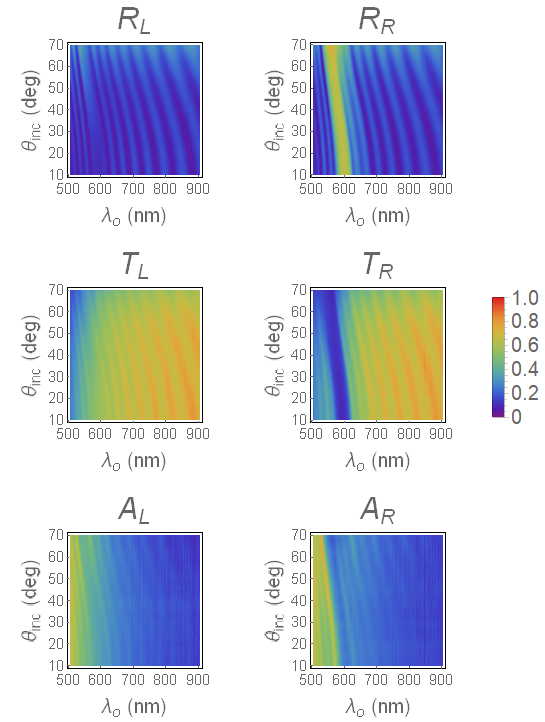}
\end{tabular}
\end{center}
\caption{\label{Expt-CircRemit}
$\RL$, $\RR$, $\TL$, $\TR$, $\AL$, and $\AR$ measured as functions
of $\lambdao$ and $\thetainc$ for $h=1$.  }
\end{figure}

Figure~\ref{Expt-CircRemit} provides measured values of
$\RL$, $\RR$, $\TL$, $\TR$, $\AL$, and $\AR$ as functions
of $\lambdao$ and $\thetainc$. Due to mechanical limitations of the optical apparatus, reflectances could only be measured for $\thetainc \geq 10$~deg, which limited the absorptance data to $\thetainc \in\lec {10\ \text{deg},70\ \text{deg}}\ric$.

The circular Bragg phenomenon is manifest in Fig.~\ref{Expt-CircRemit} as a high  ridge in the graph of $\RR$ and a correspondingly deep trough in the graph of $\TR$. 
For $\thetainc\in\left\{10\ \text{deg}, 30\ \text{deg},70\ \text{deg}\right\}$, values of the center wavelength
  of the circular Bragg regime were estimated   as  $\lambdaoBr\in\left\{596.8\ \text{nm}, 586.1\ \text{nm},551.9\ \text{nm}\right\}$ 
from both of these graphs. Also, the circular Bragg phenomenon is clearly evident as a trough in
the graphs of $\RL$ and $\AR$ and faintly as a ridge in the graphs of $\TL$ and $\AL$. Also,
 the circular Bragg regime blueshifts as the incidence becomes more oblique. All of these features are in accord with
 the data calculated for Fig.~\ref{Theor-CircRemit}.

\begin{figure}
\begin{center}
\begin{tabular}{c}
\includegraphics[width=0.55\linewidth]{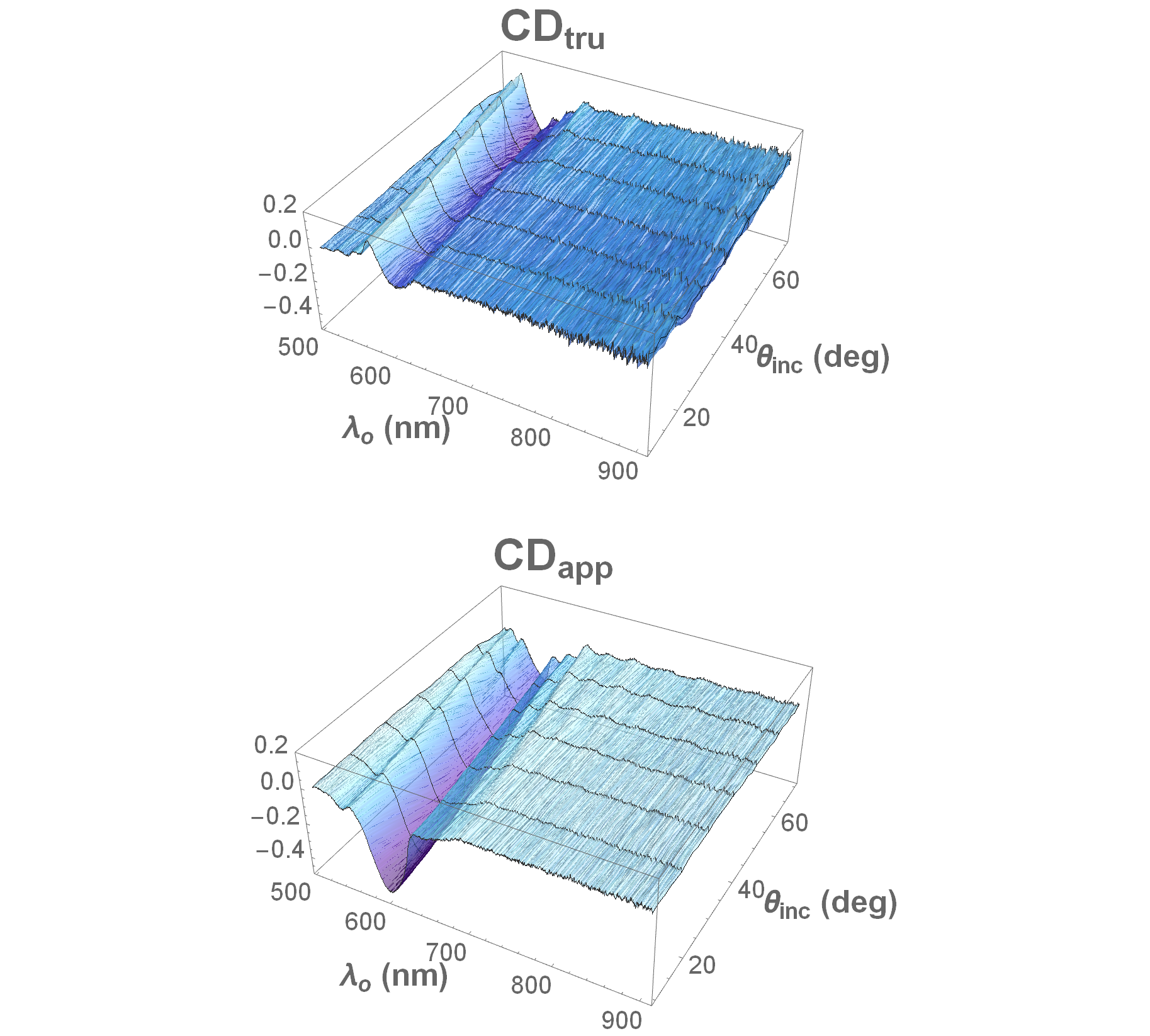}
\end{tabular}
\end{center}
\caption{\label{Expt-tCDaCD}
$\tCD$  and $\aCD$ measured as functions
of $\lambdao$ and $\thetainc$ for $h=1$ and $\psi=0$~deg. }
\end{figure}

Graphs of $\tCD$ and $\aCD$ measured as functions of $\lambdao$ and $\thetainc$ are provided
in Fig.~\ref{Expt-tCDaCD}. The trough in the graph of $\tCD$
indicative of the circular Bragg regime is less deep, of narrower bandwidth, and redshifted
in comparison to its counterpart  in the graph of $\aCD$. These features are somewhat different
than those in the theoretical graphs of Fig.~\ref{Theor-tCDaCD}, thereby indicating that
 the single-resonance Lorentzian dependences
of $\eps_{\rm a,b,c}$ on $\lambdao$ require some
modification \cite{Erten2015}.

\subsection{Linear Reflectances, Transmittances, and Dichroisms}

\begin{figure}
\begin{center}
\begin{tabular}{c}
\includegraphics[width=0.55\linewidth]{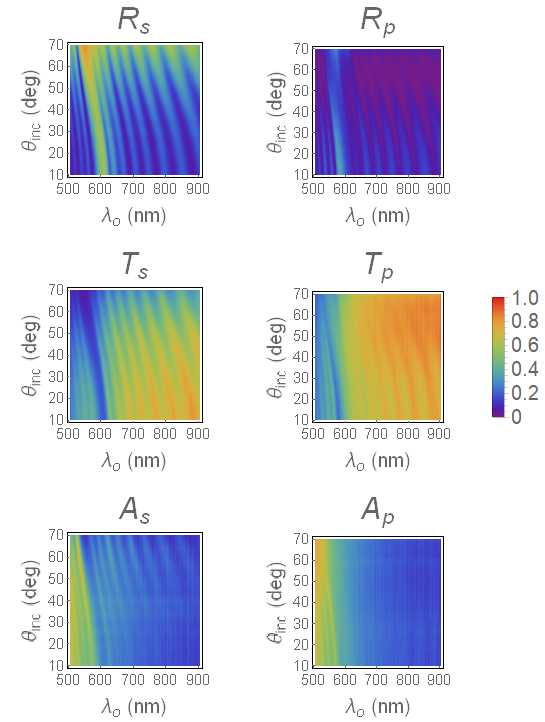}
\end{tabular}
\end{center}
\caption{\label{Expt-LinRemit}
$\Rs$, $\Rp$, $\Ts$, $\Tp$, $\As$, and $\Ap$ measured as functions
of $\lambdao$ and $\thetainc$ for $h=1$ and $\psi=0$~deg. 
}
\end{figure}

\begin{figure}
\begin{center}
\begin{tabular}{c}
\includegraphics[width=0.55\linewidth]{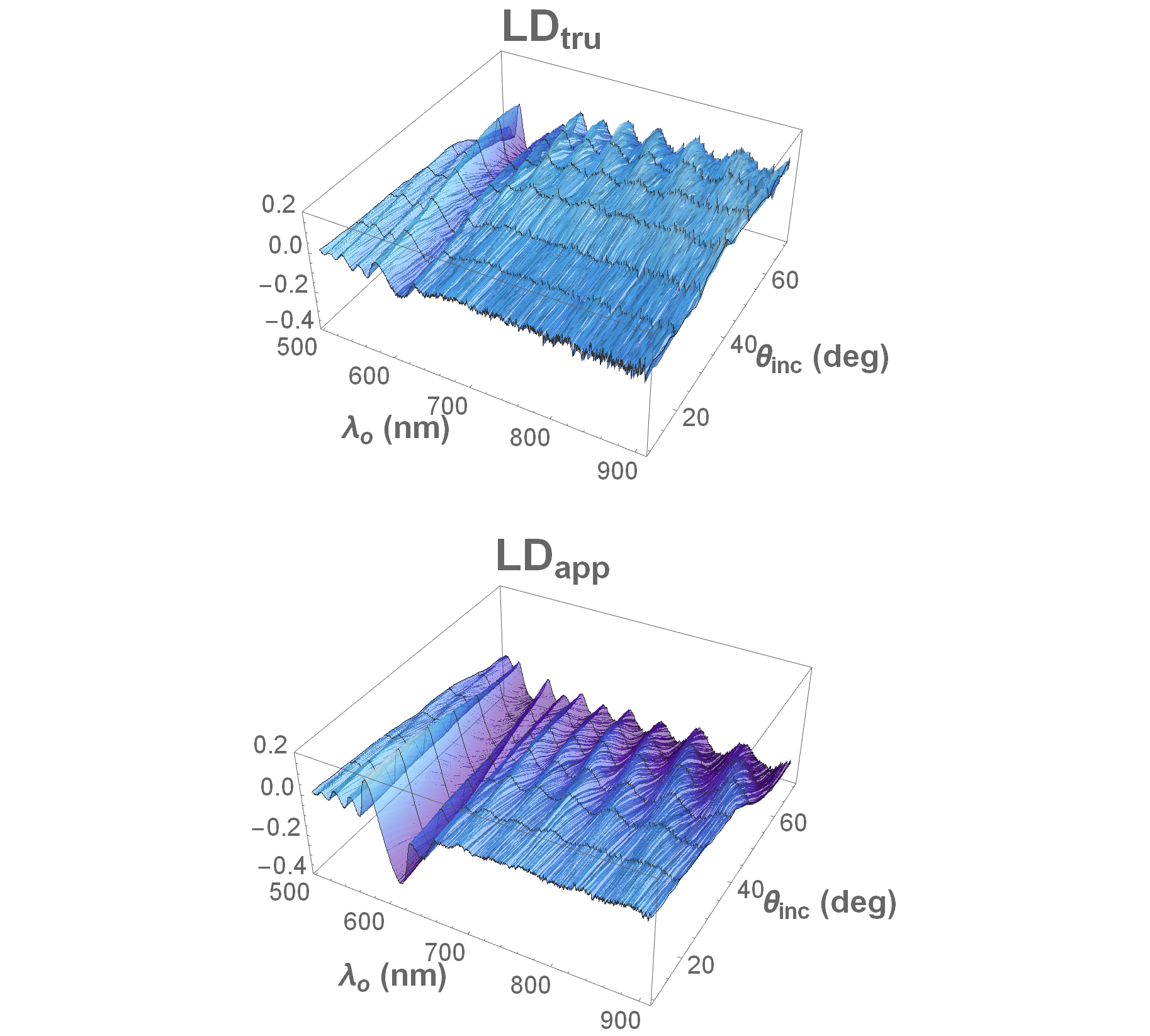}
\end{tabular}
\end{center}
\caption{\label{Expt-tLDaLD}
$\tLD$  and $\aLD$ measured as functions
of $\lambdao$ and $\thetainc$ for $h=1$ and $\psi=0$~deg. 
}
\end{figure}

Measured values of $\Rs$, $\Rp$, $\Ts$, and $\Tp$ are provided in relation
to $\lambdao$ and $\thetainc$ in Fig.~\ref{Expt-LinRemit}. In all four graphs, the circular Bragg phenomenon  
is clearly evident along with its blueshift   with oblique incidence. 

From the data for $\As$ and
$\Ap$ also depicted in Fig.~\ref{Expt-LinRemit}, we computed both $\tLD$ and $\aLD$
as functions of $\lambdao$ and $\thetainc$ for presentation in Fig.~\ref{Expt-tLDaLD}.
The circular Bragg phenomenon is manifest in all four graphs. As the value of $\psi$
could not be known, a more robust comparison with Figs.~\ref{Theor-LinRemit} and
\ref{Theor-tLDaLD} is not possible.

\section{Concluding Remarks}

We have provided both theoretical and experimental evidence for the simultaneous exhibition
of linear and circular dichroisms by a dielectric structural chiral material. Whereas both
$\tCD$ and $\tLD$ arise from dissipation in the DSCM, $\aCD$ and $\aLD$ can be exhibited
even if dissipation was weak enough to be ignored. The co-occurrence of 
both linear and circular dichroisms can only be attributed to the structural chirality of 
a DSCM,  gyrotropy being absent therein.

Our investigation covered a wide enough spectral regime to encompass the exhibition
of the circular Bragg phenomenon by the chosen DSCM. This phenomenon is manifested 
not only in the graphs of circular reflectances and transmittances but also in the graphs
of linear reflectances and transmittances. Furthermore, the circular Bragg phenomenon
is manifested in the spectrums of $\tCD$, $\tLD$, $\aCD$ and $\aLD$. The blueshift of
the circular Bragg phenomenon with increasingly oblique incidence was also captured
in the spectrums of linear reflectances and transmittances as well as in the spectrums of
circular and linear dichroisms.

We found from theory that reversing the structural handedness (i.e., $h\to-h$)  of a DSCM and 
rotating the  projection
of the direction
of propagation of the incident light  clockwise
instead of counterclockwise (i.e., $\psi\to-\psi$) about the   axis of helicoidal nonhomogeneity ($z$ axis)
simultaneously
changes the signs of both $\tCD$ and $\aCD$. However, the  simultaneous
reversals have no effect on both $\tLD$ and $\aLD$.

Before concluding, we must compare our findings with those for quartz. Quartz is a homogeneous material with a uniaxial permittivity dyadic and can display linear dichroism \cite{Ghosh}. However, even in the early 1800s,  natural optical activity was observed when both plane waves propagate parallel to the sole optic axis of the material, as was remarked upon by Condon \cite{Condon}. As has been  inferred only very recently by experimentation \cite{Nichols}, quartz has to be characterized as a
homogeneous bianisotropic material \cite{ML-EAB}, whose permittivity and both magnetoelectric dyadics are uniaxial and whose permeability dyadic is scalar. In contrast, the DSCMs considered in our work are purely dielectric materials without gyrotropy but are nonhomogeneous along a fixed direction \cite{deG,Nityananda,YK1959,STFbook,Erten2015}.




\end{document}